\documentclass[%
 reprint,superscriptaddress,nolongbibliography,
 amsmath,amssymb,
 aps,
 floatfix,
]{revtex4-2}

\usepackage{graphicx}
\usepackage{capt-of}
\usepackage{dcolumn}
\usepackage{bm}
\usepackage{natbib}
\usepackage[colorlinks=true,linkcolor=blue,citecolor=blue,urlcolor=blue]{hyperref}

\begin{document}

\title{Electric Field Induced Multi-Space Topological Phase Transitions in Janus Monolayer MnBi$_2$Se$_2$Te$_2$}

% 第一组：清华大学（第一作者 + 第二作者）
\author{Kefan Zhang}
\author{Jian Wu}
\affiliation{
    State Key Laboratory of Low Dimensional Quantum Physics,
    Department of Physics,
    Tsinghua University, Beijing 100084, China
}
% 第二组：雷根斯堡大学（通讯作者，带星号和邮箱）
\author{Weiyi Pan}
\email{Weiyi.Pan@physik.uni-regensburg.de}
%\thanks{Corresponding author}
\affiliation{
    Institute for Theoretical Physics,
    University of Regensburg, 93040 Regensburg, Germany
}

\date{\today}

\begin{abstract}
Manipulating coexisting multi-space topological states within a single material is a critical frontier for low-power spintronic applications, for which perpendicular electric-field gating offers a highly precise tuning mechanism. Using first-principles calculations and atomistic spin simulations, it is demonstrated that Janus monolayer MnBi$_2$Se$_2$Te$_2$ can exhibit simultaneous momentum-space and real-space topological phase transitions under an external electric field. Specifically, the electric field drives momentum-space transitions across a topologically trivial state ($C = 0$), a low-Chern-number quantum anomalous Hall state ($C = -1$), and a high-Chern-number state ($C = 2$). Concurrently, by actively tuning the competition between the Dzyaloshinskii-Moriya interaction and magnetic anisotropy, the electric field induces real-space magnetic transitions from a uniform ferromagnetic state to an isolated skyrmion state, and ultimately to a skyrmion-spiral domain coexistence phase.  Electric-field-induced variations in both the QAHE and magnetic textures may give rise to multiple topological phase transitions, involving distinct topological regimes: a purely k-space topology, RK-joint skyrmions, and pure real-space skyrmions. These findings establish a powerful material platform and efficient route for the synergistic control of coexisting topological orders, making it highly promising for next generation multi-functional spintronic devices. 
\end{abstract}

\maketitle

\section{Introduction}

Topology has fundamentally revolutionized condensed matter physics, catalyzing the discovery of numerous novel quantum states \cite{kosterlitz1973ordering,thouless1982quantized,hasan2010colloquium,qi2011topological}. In magnetic materials and heterostructures, topological orders can coexist in both momentum space ($k$-space) and real space. Specifically, k-space topology manifests primarily as electronic band topology driven by spin-orbit coupling (SOC), whereas real-space topology corresponds to topologically non-trivial magnetic spin textures formed by non-collinear spin arrangements \cite{hasan2010colloquium,nagaosa2013topological,tokura2020magnetic}. A key challenge is to realize the coexistence and coupling of k-space and real-space topology within a single material system, and to manipulate them in a controlled manner. Achieving this goal would not only deepen our understanding of multiple topological phenomena, but also open opportunities for future multifunctional devices \cite{fert2017magnetic,he2022topological,han2025electric}.

A hallmark of k-space topology is the quantum anomalous Hall (QAH) effect, which features dissipationless chiral edge states at zero magnetic field \cite{haldane1988model,weng2015quantum,liu2016quantum}. It arises when intrinsic magnetization and strong SOC induce a topological band inversion, yielding a quantized anomalous Hall conductivity (AHC) governed by the Chern number. The magnitude and sign of the Chern number determine the number and chirality of chiral edge channels, respectively, so controllable tuning of the Chern number is essential for reconfigurable dispassionless spintronic devices. Traditionally, adjusting chirality relies on magnetization reversal \cite{chang2013experimental,xu2022controllable}, whereas modifying the magnitude usually requires altering material thickness \cite{zhao2020tuning,deng2020quantum}, applying mechanical strain \cite{chen2024strain,li2022interplay,yang2023strain}, or constructing moir\'{e} superlattices \cite{serlin2020intrinsic}. Electric-field control of Berry curvature and Chern insulating phases has also been demonstrated in related 2D magnetic systems \cite{du2020berry,you2021electric}. However, simultaneously controlling both parameters within a single platform remains a formidable challenge. More critically, executing such comprehensive modulation exclusively via a perpendicular electric field--a reversible and localized mechanism essential for integrated devices--deserves further investigation in realistic material systems.

Beyond the above-mentioned k-space electronic topology, real-space topology can emerge in magnetic systems in the form of magnetic skyrmions, which are topologically protected non-collinear spin textures \cite{nagaosa2013topological,zhang2020skyrmion,tokura2017emergent,tokura2020magnetic}. These topologically protected spin textures are driven by the competition between the Dzyaloshinskii-Moriya interaction (DMI) and Heisenberg exchange interaction \cite{du2022spontaneous,pang2023electric,wei2026tunable}. For practical device applications of skyrmions, reliable writing and erasing operations are essential \cite{fert2013skyrmions,sampaio2013nucleation,romming2013writing}. Conventionally, these operations can be achieved by modulating skyrmion stability through an externally applied electric field in magnetic materials. However, such electric-field control has so far been realized and observed primarily in magnetic heterostructures \cite{paul2022electric,desplat2021mechanism,li2023tuning}, in which the existence of interfacial defects would pin the skyrmions \cite{fernandes2020impurity,fernandes2018universality} and thus hinders their controlled motion and manipulation. Consequently, achieving purely electric-field-driven skyrmion manipulation within an intrinsic, single-material platform is highly desirable. Furthermore, skyrmions have been reported to couple with k-space electronic topology, potentially giving rise to novel topological phenomena and multifunctional quantum devices \cite{jiang2020concurrence,zhang2018real,li2022interplay}. However, simultaneously controlling these two topological orders within a single material system, especially through purely electrical means, remains challenging.

In this work, we propose a simultaneous all-electric control of both the QAH effect and magnetic skyrmions in Janus topological magnet MnBi$_2$Se$_2$Te$_2$ (MBTSe). Based on first-principles calculations and atomistic spin dynamic simulations, it is revealed that a perpendicular electric field simultaneously drives topological transitions in both k-space and real space. In k-space, the system undergoes successive transitions among a topologically trivial state (C = 0), a low-Chern-number QAH state (C = -1), and a high-Chern-number QAH state (C = 2) through bandgap closing and reopening under electric field. In real space, the electric field induces magnetic transitions from a uniform ferromagnetic (FM) state to a skyrmion phase, which is also revealed by the variation of the skyrmion collapse barrier. This coupled evolution leads to the formation of multi-space joint topological structures. These findings highlight the convenience and efficiency of electric-field gating in controlling multiple topological orders, paving the way for the design of multi-functional topological spintronic devices.

\section{Computational Methods}

\subsection{First-Principles Calculations}

First-principles calculations are performed using the Vienna Ab initio Simulation Package (VASP) \cite{kresse1996efficient}. The projector augmented-wave (PAW) method \cite{blochl1994projector} is employed with a plane-wave energy cutoff of 400 eV. To eliminate periodic image interactions, a 15 \AA{} vacuum layer is introduced. The Perdew-Burke-Ernzerhof (PBE) functional is utilized for the exchange-correlation functional in the Kohn-Sham equations \cite{perdew1996generalized}. The semi-core states of Mn-$3d$ and Bi-$5p$ orbitals are treated as valence electrons. To describe the strong correlation effects of Mn $3d$ electrons, the DFT$+U$ method \cite{dudarev1998electron} is adopted with U = 4 eV for Mn \cite{li2019intrinsic,zhang2021tunable}. Both the lattice constants and the relative coordinates of all atoms are fully relaxed until the residual forces are below 0.01 eV/\AA{}. A 18 $\times$ 18 $\times$ 1 Monkhorst-Pack k-point mesh is used to ensure the accuracy of our calculations. The total energy convergence criteria are set to $10^{-7}$ eV for structural relaxation and $10^{-8}$ eV for static self-consistent calculations. To account for the influence of an external electric field on this polar system, a dipole correction is applied to obtain more reliable results.

\subsection{Magnetic Interaction}

The magnetic interactions of the localized Mn-$3d$ electrons can be described by the following spin Hamiltonian:

\begin{equation}
\mathcal{H} = \sum_{i,j} J_{ij} \mathbf{S}_i \cdot \mathbf{S}_j + \sum_{i,j} \mathbf{D}_{ij} \cdot (\mathbf{S}_i \times \mathbf{S}_j) + A_{zz} \sum_i (S_i^z)^2
\end{equation}

Here, $J_{ij}$ is the Heisenberg exchange interaction, with $J_{ij}<0$ and $J_{ij}>0$ denoting FM and antiferromagnetic (AFM) coupling, respectively. Three nearest-neighbor exchange interactions ($J_1$, $J_2$, and $J_3$) are considered; the detailed calculation method for $J$ is provided in the Supplementary Material (SM) \cite{supplemental}. The MAE is defined as $A_{zz} = (E_z - E_x) / S^2$, where $E_z$ and $E_x$ are the total energies with spins aligned along the out-of-plane $z$ axis and in-plane $x$ axis, respectively. Thus, $A_{zz}<0$ indicates an out-of-plane easy axis. The DMI vector $\mathbf{D}_{ij}$ between Mn atoms $i$ and $j$ can be written as $\mathbf{D}_{ij}=d_z\mathbf{e}_z+d_{//}(\mathbf{e}_{ij}\times\mathbf{e}_z)$, where $d_{//}$ and $d_z$ are the in-plane and out-of-plane DMI components, respectively. For two-dimensional triangular magnetic systems with $C_{3v}$ symmetry, previous studies have shown that the out-of-plane component $d_z$ has a negligible influence on the stabilization of spin spirals and skyrmions \cite{moriya1960anisotropic,niu2024reducing}. We therefore focus on the in-plane component $d_{//}$, which is extracted from the energy difference between clockwise (CW) and anti-clockwise (ACW) chiral spin configurations (see the SM for details):

\begin{equation}
d_{//} = \frac{E_{\mathrm{ACW}} - E_{\mathrm{CW}}}{12 S^2}
\end{equation}

where $E_{\mathrm{ACW}}$ and $E_{\mathrm{CW}}$ denote the energies of the MBTSe monolayer with CW and ACW spin configurations, respectively, and $S$ represents the normalized spin magnetic moment (i.e., $S=1$).

\subsection{Wannier Functions and Edge States}

To investigate the topological properties, maximally localized Wannier functions (MLWFs) \cite{marzari2012maximally} are constructed using wannier90 \cite{mostofi2014updated}, with Bi-$p$, Mn-$d$, Se-$p$, and Te-$p$ orbitals included in the projection. Based on the effective Hamiltonian matrix, the electronic band structure is fitted using the Wannier function interpolation approach. In postw90, the Berry curvature is evaluated from the generalized Kubo formula \cite{wang2006ab,yates2007spectral}:
\begin{equation}
\Omega_z(\mathbf{k})=-2\hbar^2\,\mathrm{Im}\sum_{n}^{\mathrm{occ}}\sum_{m\ne n}\frac{\langle u_{n\mathbf{k}}|\hat{v}_x|u_{m\mathbf{k}}\rangle\langle u_{m\mathbf{k}}|\hat{v}_y|u_{n\mathbf{k}}\rangle}{(\varepsilon_{m\mathbf{k}}-\varepsilon_{n\mathbf{k}})^2},
\end{equation}
where $\varepsilon_{n\mathbf{k}}$ and $\hat{v}_{x/y}$ denote the band eigenvalue and velocity operator, respectively. Edge states are calculated by iteratively solving the Green's function for a semi-infinite system using the WANNIERTOOLS package \cite{sancho1985highly,wu2018wanniertools}.

\begin{figure*}[t]
\centering
\includegraphics[width=0.9\textwidth]{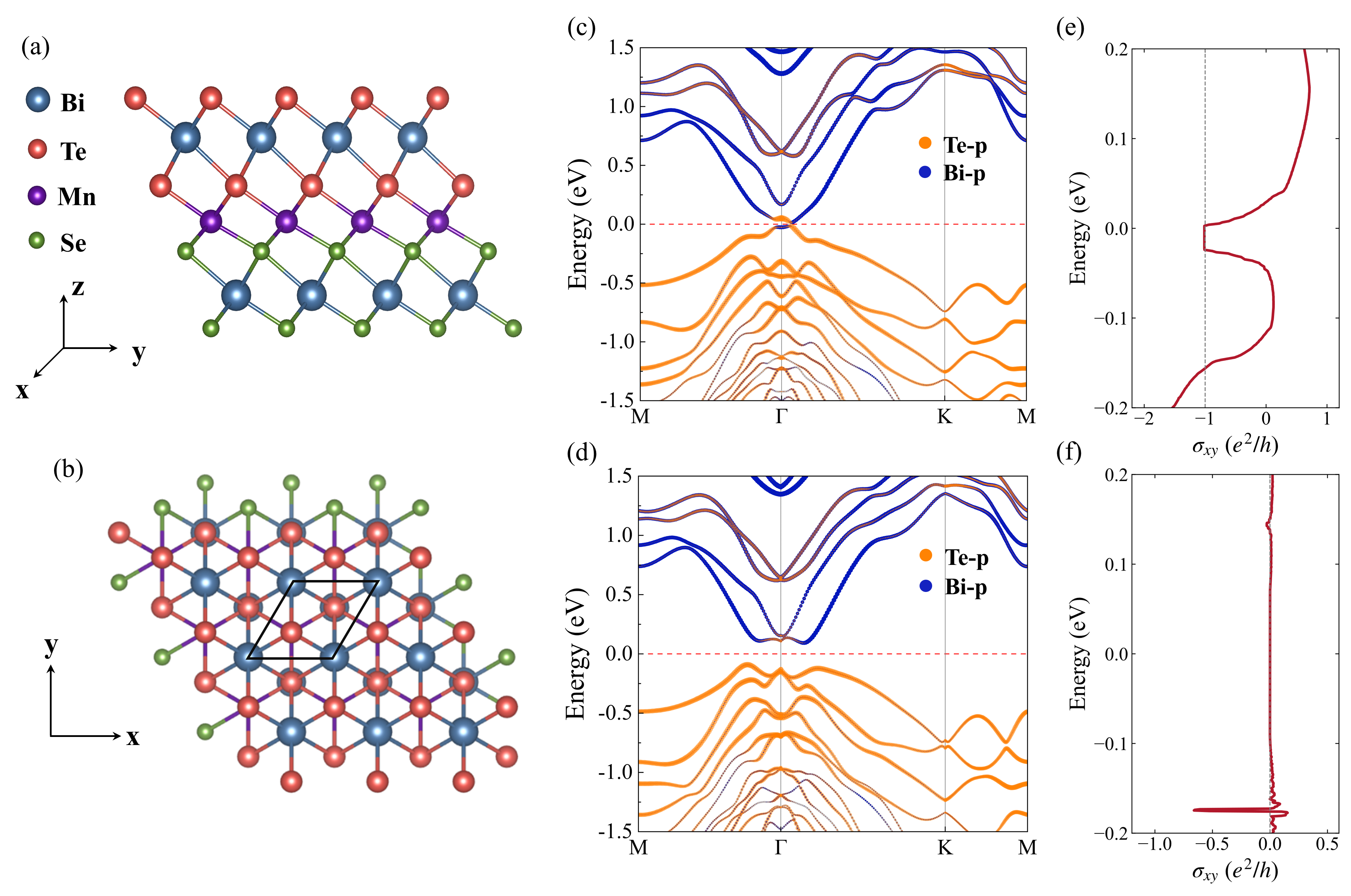}
\caption{(a) Side view of the MBTSe crystal structure and constituent atomic species. (b) Top view of the MBTSe lattice structure with unit cell labeled. (c), (e) Spin-projected band structure and the corresponding AHC for the out-of-plane FM configuration (spins aligned along the $z$ axis), respectively. (d), (f) Spin-projected band structure and the corresponding AHC for the in-plane FM configuration (spins aligned along the $x$ axis), respectively.}
\label{fig:basic_properties}
\end{figure*}

\subsection{LLG Simulations and GNEB Calculations}

To obtain the real-space magnetic configurations, atomistic spin dynamics simulations are performed by solving the Landau-Lifshitz-Gilbert (LLG) equation as implemented in the Spirit code \cite{muller2019spirit}. A $120 \times 120$ supercell is used, and the Gilbert damping parameter is set to 0.2. The time step for each LLG step is $4 \times 10^{-4}$ ps. Convergence is reached when the residual effective torque falls below $10^{-7}$ eV/\AA{}. All presented magnetic configurations successfully reach the final convergence criterion.

The stability of isolated skyrmions (ISK) is then evaluated using the geodesic nudged elastic band (GNEB) method \cite{bessarab2015method,muller2018duplication,paul2022electric}, which searches for the minimum energy path (MEP) connecting the skyrmion state and the FM state. The collapse energy barrier is extracted from this MEP, with the same residual effective torque convergence threshold of $10^{-8}$ eV/\AA{} used in the GNEB calculations.

\section{Results and Discussion}

\subsection{Basic Properties of ML MBTSe}

\begin{figure*}[t]
\centering
\includegraphics[width=0.85\textwidth]{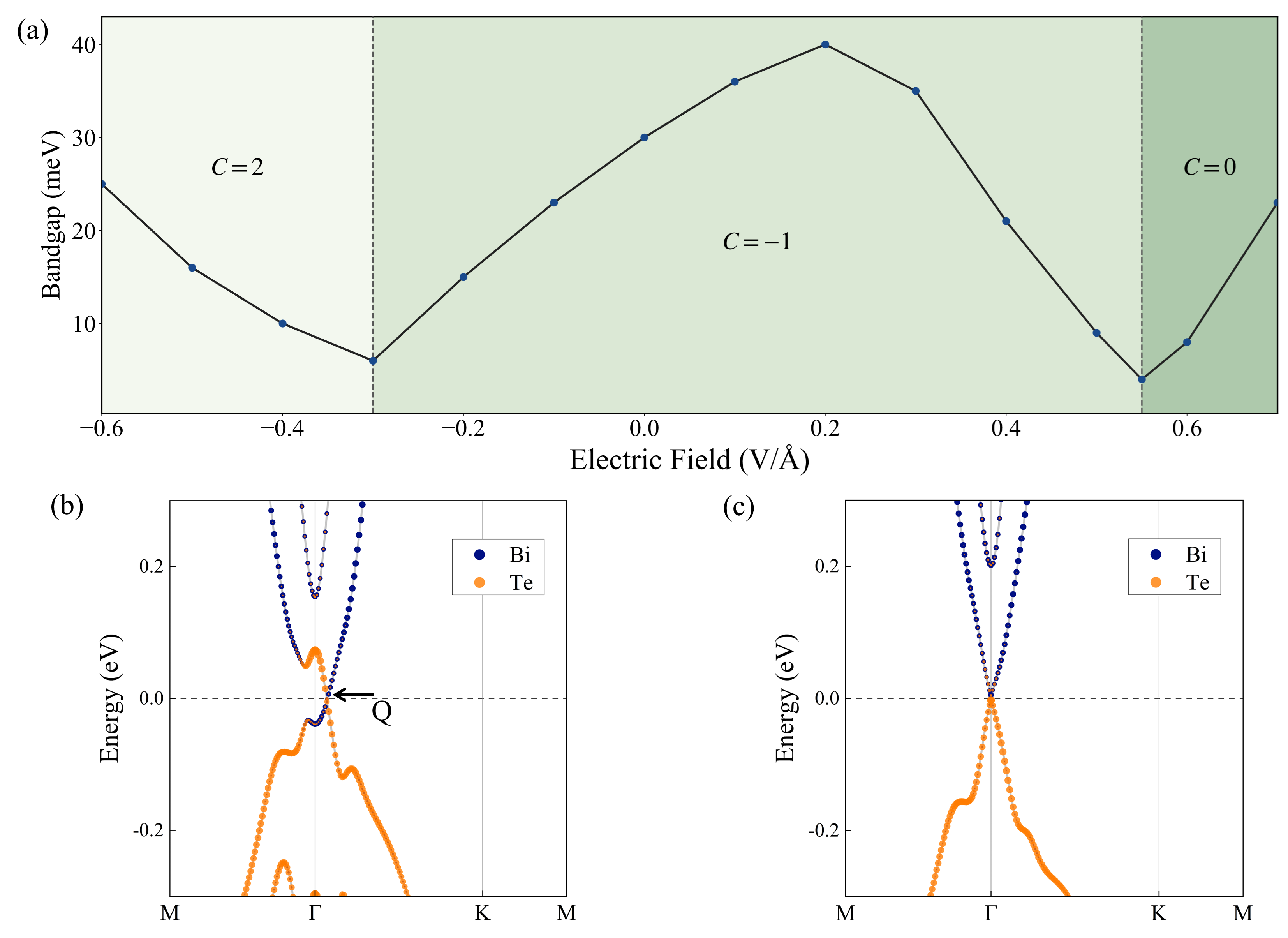}
\caption{(a) Evolution of the local bandgap as a function of the perpendicular external electric field for the out-of-plane spin configuration, revealing two critical gap-narrowing regimes. (b), (c) Enlarged orbital-projected band structures near the Fermi level for the out-of-plane spin configuration under applied electric fields of (b) $-0.3$ V/\AA{} and (c) $0.6$ V/\AA{}. These panels explicitly illustrate the field-induced band anti-crossings at the $Q$ point and the $\Gamma$ point, respectively.}
\label{fig:field_band_evolution}
\end{figure*}

The lattice structure of monolayer MBTSe is illustrated in Fig.~\hyperref[fig:basic_properties]{1(a)}. MBTSe is a Janus derivative of monolayer MnBi$_2$Te$_4$ (MBT), constructed by selective Se substitution on one side of the septuple layer. The optimized lattice constant is 4.24 \AA{}, smaller than that of MBT (4.36 \AA{}). Its most energetically stable configuration with both dynamical and thermal stability is formed by replacing the Te atoms in the Te-Bi-Te trilayer on one side of MBT with Se atoms \cite{li2022interplay}, producing a Te-Bi-Te-Mn-Se-Bi-Se stacking sequence from top to bottom. This Janus stacking breaks inversion symmetry and gives the monolayer a noncentrosymmetric P3m1 structure with a C$_3$ rotation along the z-axis and several mirror symmetries [Fig.~\hyperref[fig:basic_properties]{1(b)}]. Consequently, such specific lattice structure of MBTSe naturally induces a strong DMI, providing an ideal platform for the generation of real-space non-trivial topological phases, such as skyrmions. The Mn atoms are the primary source of magnetism. They are located at the centers of distorted octahedra and are arranged in a two-dimensional triangular lattice. DFT calculations show that the Mn-3$d$ orbitals adopt a high-spin $d^5\uparrow d^0\downarrow$ configuration, leading to a total spin magnetic moment of $S = 5/2$ for each Mn atom.

The parameters of the spin Hamiltonian Eq.~(1) are calculated using the energy-mapping methods. The dominant nearest-neighbor exchange interaction $J_1 = -1.493$ meV, which strongly favors local FM rather than AFM coupling between adjacent spins in MBTSe. The system exhibits an in-plane DMI of $d_{//} = 0.284$ meV and an out-of-plane MAE of $A_{zz} = 0.098$ meV.  Crucially, the resulting DMI-to-exchange ratio ($d_{//}/|J_1| \approx 0.19$) lies within the commonly reported range of $0.1$--$0.2$ that is favorable for stabilizing skyrmionic phases in two-dimensional magnetic systems \cite{liang2020very,du2022spontaneous,zhang2023generation}.

When MBTSe is in the out-of-plane FM configuration, its element-projected band structure is shown in Fig.~\hyperref[fig:basic_properties]{1(c)}, with the corresponding AHC presented in Fig.~\hyperref[fig:basic_properties]{1(e)}. The states near the Fermi level are mainly contributed by Bi-$p$ and Te-$p$ orbitals, whereas the Mn-$d$ bands lie far from the Fermi level due to the strong correlation of the localized Mn 3$d$ electrons (see the full element-projected band structure in the SM). After SOC is included, a band inversion between Bi-$p$ and Te-$p$ orbitals appears near the $\Gamma$ point and opens a gap of about 29 meV. Such a characteristic band inversion strongly indicates the emergence of a non-trivial electronic band topology. The calculated AHC is quantized, demonstrating that this out-of-plane FM state is a QAH insulator with $C = -1$. Conversely, when the symstem is in the in-plane FM configuration, the band structures for spins aligned along both the $x$ and $y$ directions exhibit significantly larger bandgaps of approximately 180 meV near the Fermi level, without obvious anti-crossings [Fig.~\hyperref[fig:basic_properties]{1(d)}]. The calculated AHC becomes zero for these in-plane magnetic configurations, indicating a topologically trivial indirect-gap semiconductor [Fig.~\hyperref[fig:basic_properties]{1(f)}]. 

\subsection{Band Topology with External Electric Field}

\begin{figure*}[t]
\centering
\includegraphics[width=\textwidth]{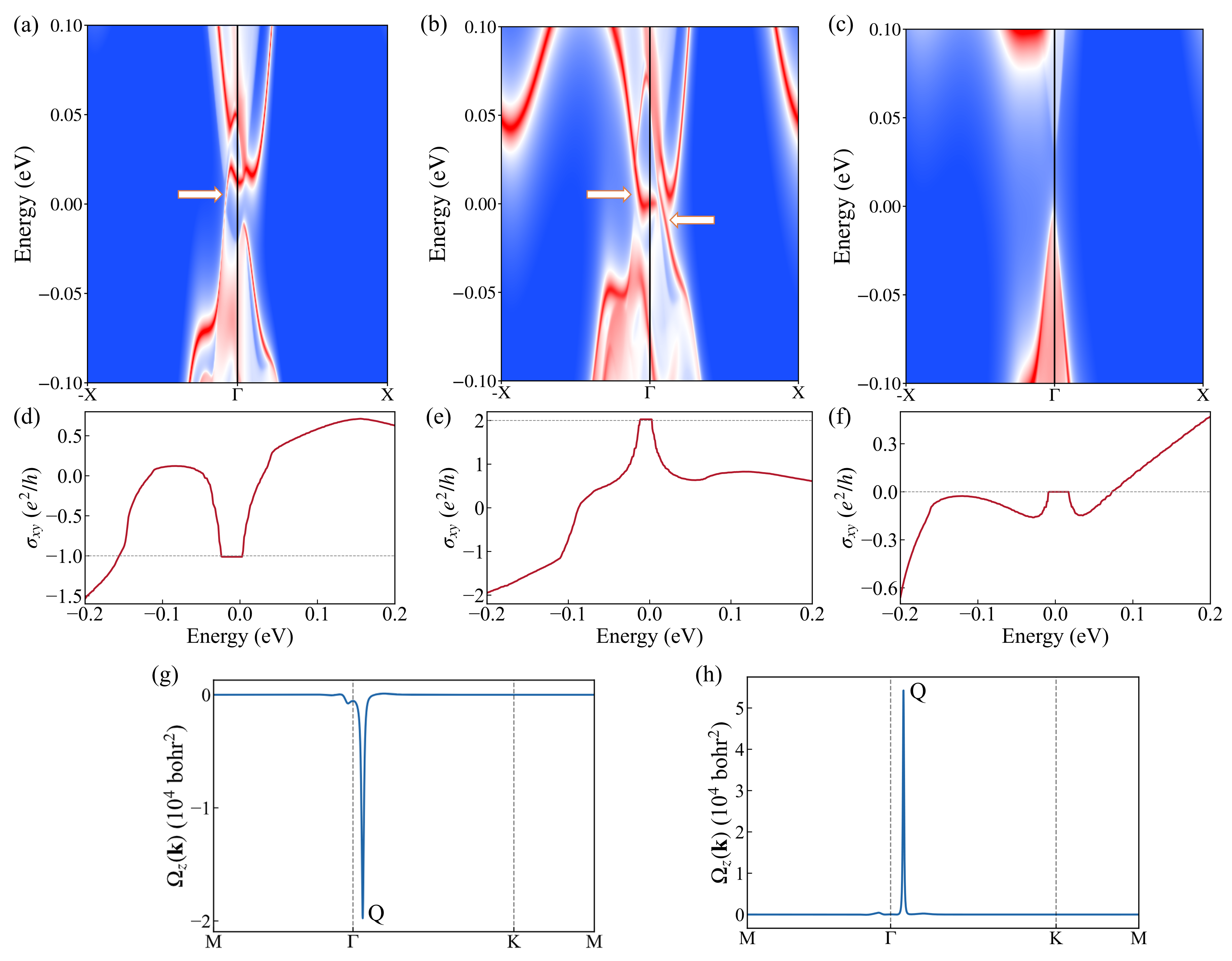}
\caption{Calculated chiral edge states for semi-infinite MBTSe at perpendicular electric fields of (a) $E = 0$ V/\AA{}, (b) $E = -0.5$ V/\AA{}, and (c) $E = 0.6$ V/\AA{}. Arrows in the panels indicate the positions of the chiral edge states. (d)-(f) Corresponding AHC as a function of Fermi energy, explicitly demonstrating topological phases with Chern numbers $C = -1$, $C = 2$, and $C = 0$, respectively. (g)-(h) Berry curvature $\Omega_z(\mathbf{k})$ plotted along high-symmetry lines at (g) $E = 0$ V/\AA{} and (h) $E = -0.5$ V/\AA{}, highlighting the pronounced response at the Q point associated with the phase transitions.}
\label{fig:topological_phases}
\end{figure*}

Following the characterization of the fundamental electronic and magnetic properties of MBTSe, the subsequent analysis focuses on how an external electric field modifies its electronic structure and band topology. To this end, band structures are computed under varying perpendicular electric fields, and the evolution of the local bandgap is systematically tracked. As shown in Fig.~\hyperref[fig:field_band_evolution]{2(a)}, the bandgap exhibits two critical crossover regimes as the electric field is varied. Under a negative electric field, the gap continuously narrows and reaches a minimum near $-0.3$ V/\AA{}; under a positive electric field, a second gap-narrowing process occurs near $0.55$ V/\AA{}. The enlarged projected band structures near the Fermi level demonstrate that these gap-closing events are driven by field-induced band crossings located at threefold Q points and at the $\Gamma$ point, respectively [Fig.~\hyperref[fig:field_band_evolution]{2(b)}], with the broader electric-field-dependent near-Fermi band evolution shown in the SM \cite{supplemental}.

To reveal the topological nature of these crossovers, Wannier-function models are established to calculate the AHC and chiral edge states before and after the critical fields. As shown in Fig.~\hyperref[fig:topological_phases]{3(a)-(c)}, the system transitions through three distinct topological phases. For the pristine and low-field regime, the system is a QAH insulator with a Chern number of $C = -1$, hosting one chiral edge state [Fig.~\hyperref[fig:topological_phases]{3(b)}]. When the negative electric field exceeds $-0.3$ V/\AA{}, the Chern number transitions to $C = 2$, accompanied by the emergence of two chiral edge states with opposite chirality [Fig.~\hyperref[fig:topological_phases]{3(a)}]. Conversely, when the electric field is larger than $+0.55$ V/\AA{}, the Chern number is neutralized to $C = 0$, and the chiral edge states completely disappear, transforming the system into a topologically trivial semiconductor [Fig.~\hyperref[fig:topological_phases]{3(c)}].

\begin{figure*}[t]
\centering
\includegraphics[width=\textwidth]{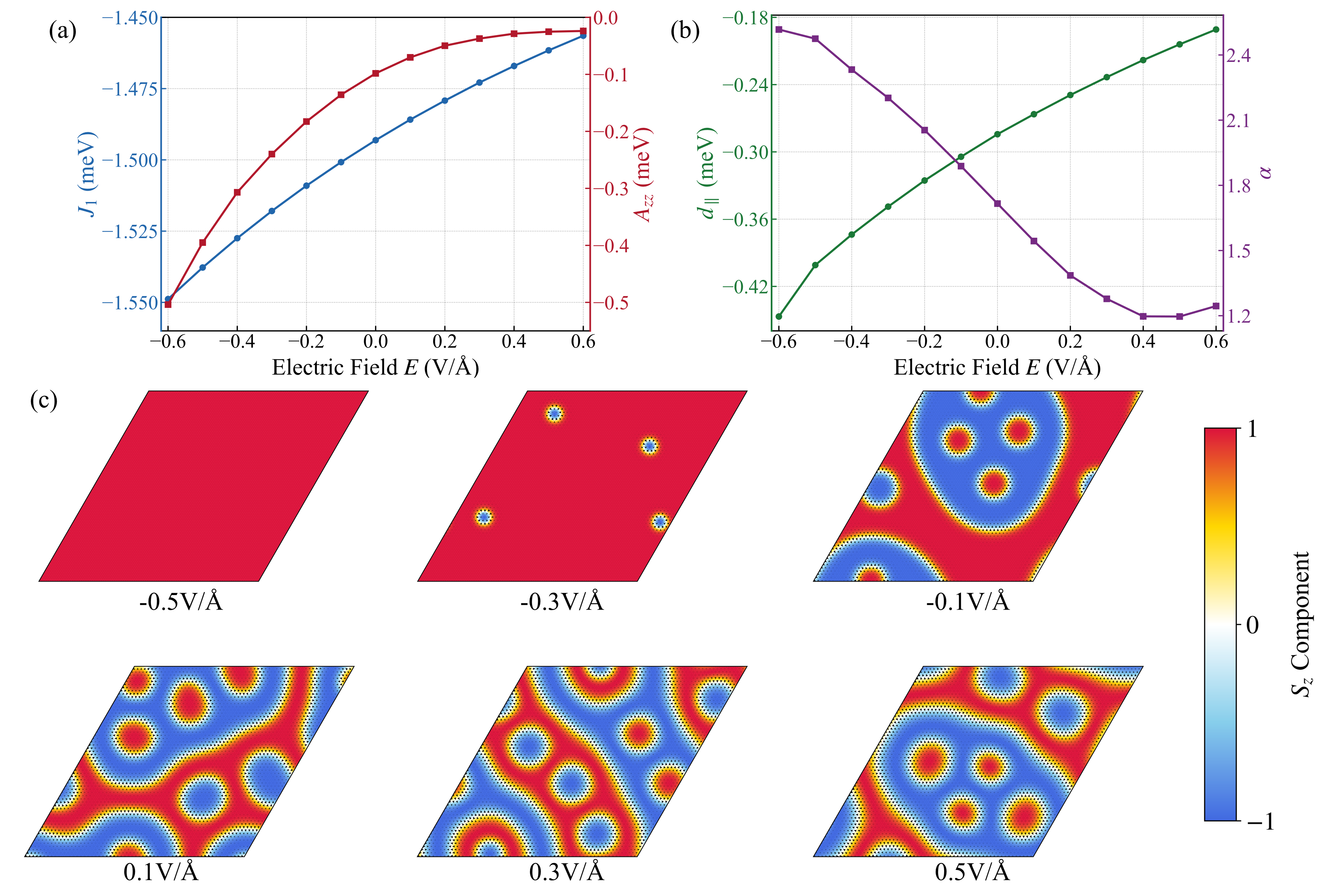}
\caption{(a) Evolution of the nearest-neighbor exchange interaction $J_1$ and the MAE $A_{zz}$ as a function of the external electric field $E$. (b) Electric-field dependence of the in-plane DMI $d_{//}$ and the empirical skyrmion-stabilization parameter $\alpha$. (c) Atomistic LLG simulations of real-space magnetic configurations under varying electric fields, illustrating the transition from a uniform FM state to ISK, and ultimately to a skyrmion-spiral domain (SK-SD) coexistence phase.}
\label{fig:magnetic_configuration}
\end{figure*}

The physical mechanism driving these topological phase transitions was subsequently investigated by tracking how the electric-field-induced electrostatic potential shifts tune the relative band energies, leading to band anti-crossings and the associated Berry curvature redistribution. Under a negative electric field, the critical anti-crossings occur at the Q points. Due to the $C_{3z}$ symmetry of the lattice, three equivalent Q points in the Brillouin zone undergo simultaneous anti-crossings, generating three new chiral edge states and driving the Chern number from $-1$ to $2$. In contrast, under a positive electric field, the critical anti-crossing is confined to the single $\Gamma$ point at the zone center, inducing a precise Chern number change of $+1$ and shifting the system from $-1$ to $0$. This mechanism is further supported by the calculated Berry curvature distributions of the two topologically non-trivial phases, as shown in Fig.~\hyperref[fig:topological_phases]{3(d)}; the detailed Berry curvature distributions in the first Brillouin zone are provided in the SM \cite{supplemental}. Compared with the $C = -1$ phase, the $C = 2$ phase induced by the negative electric field exhibits pronounced sign reversal and intensification of the local Berry curvature around the three symmetry-equivalent Q points, where the band anti-crossings occur. This Q-point-resolved Berry curvature response is therefore fully consistent with the observed Chern number transition from $-1$ to $2$.

\subsection{External Field Tunable Magnetic Configuration of ML MBTSe}

After confirming the topological phase transitions in momentum space, attention is turned to real space. Investigating parallel transitions in magnetic configurations is highly significant for demonstrating the multi-space topological nature of the system. The magnetic interaction parameters $J$, $d_{//}$, and $A_{zz}$ under different electric fields were calculated, and the specific results are presented in Fig.~\hyperref[fig:magnetic_configuration]{4(a)}. The dominant FM nearest-neighbor interaction $J_1$ ($\sim -1.5$ meV) remains relatively stable under the electric field, rendering the much smaller $J_2$ and $J_3$ (detailed in the SM) magnetically negligible.

The DMI strength ranges between 0.2 and 0.45 meV, and its value varies significantly with the electric field. A negative electric field enhances the DMI value, whereas a positive electric field suppresses it. The MAE values range between 0.02 and 0.5 meV, exhibiting an out-of-plane magnetic easy axis across all evaluated electric field ranges. The trend of the MAE variation matches that of the DMI strength, but its amplitude of change is noticeably larger. To evaluate the tendency of MBTSe to form topologically non-trivial skyrmion configurations, an empirical parameter $\alpha = \frac{4\sqrt{J_1 A_{zz}}}{\pi d_{//}}$ was calculated, where a smaller value indicates a higher propensity for skyrmion generation \cite{banerjee2014enhanced,soumyanarayanan2017tunable}. As the magnitude of the negative electric field increases, the enhancement of MAE is significantly more pronounced than that of DMI. Consequently, the magnetic moments are rigidly confined to the out-of-plane direction, which drives the system toward a topologically trivial uniform FM state and severely suppresses the generation of skyrmions. Conversely, as the positive electric field increases, the situation is reversed; the rapid reduction in MAE relieves the out-of-plane spin confinement, making the stabilization of skyrmions energetically favorable. However, a notable exception emerges when the positive electric field exceeds $0.5$ V/\AA{}. In this high-field regime, the approach of $A_{zz}$ toward zero and the concurrent reduction of $d_{//}$ jointly govern the evolution of the empirical parameter $\alpha$; their combined effect leads to an increase in $\alpha$, making the conditions unfavorable for skyrmion stabilization.

The LLG simulations of the magnetic configurations are consistent with these assessments. When the electric field ranges from $-0.6$ V/\AA{} to $-0.4$ V/\AA{}, the magnetic configuration of the system is FM. As the electric field increases (becomes more positive), ISK begin to appear, and their radii gradually increase. When the electric field reaches $-0.1$ V/\AA{}, the LLG simulations reveal a magnetic configuration where the skyrmion phase coexists with spiral domain (SK-SD). This transitional trend from FM to ISK and then to SK-SD demonstrates that as the electric field becomes more positive, the system progressively favors the generation of skyrmions, as further illustrated by the detailed electric-field-dependent magnetic configurations in the SM \cite{supplemental}. Notably, although the empirical parameter $\alpha$ indicates a theoretically less favorable condition at extreme positive fields (exceeding $0.5$ V/\AA{}), the LLG simulations confirm that this high-field anomaly does not destabilize the established SK-SD coexistence phase, which robustly persists. Therefore, an external electric field can drive the real-space topology of MBTSe by modulating the system's magnetic interactions, particularly the strengths of the DMI and MAE.

\begin{figure}[b!]
%\centering
\includegraphics[width=0.95\columnwidth]{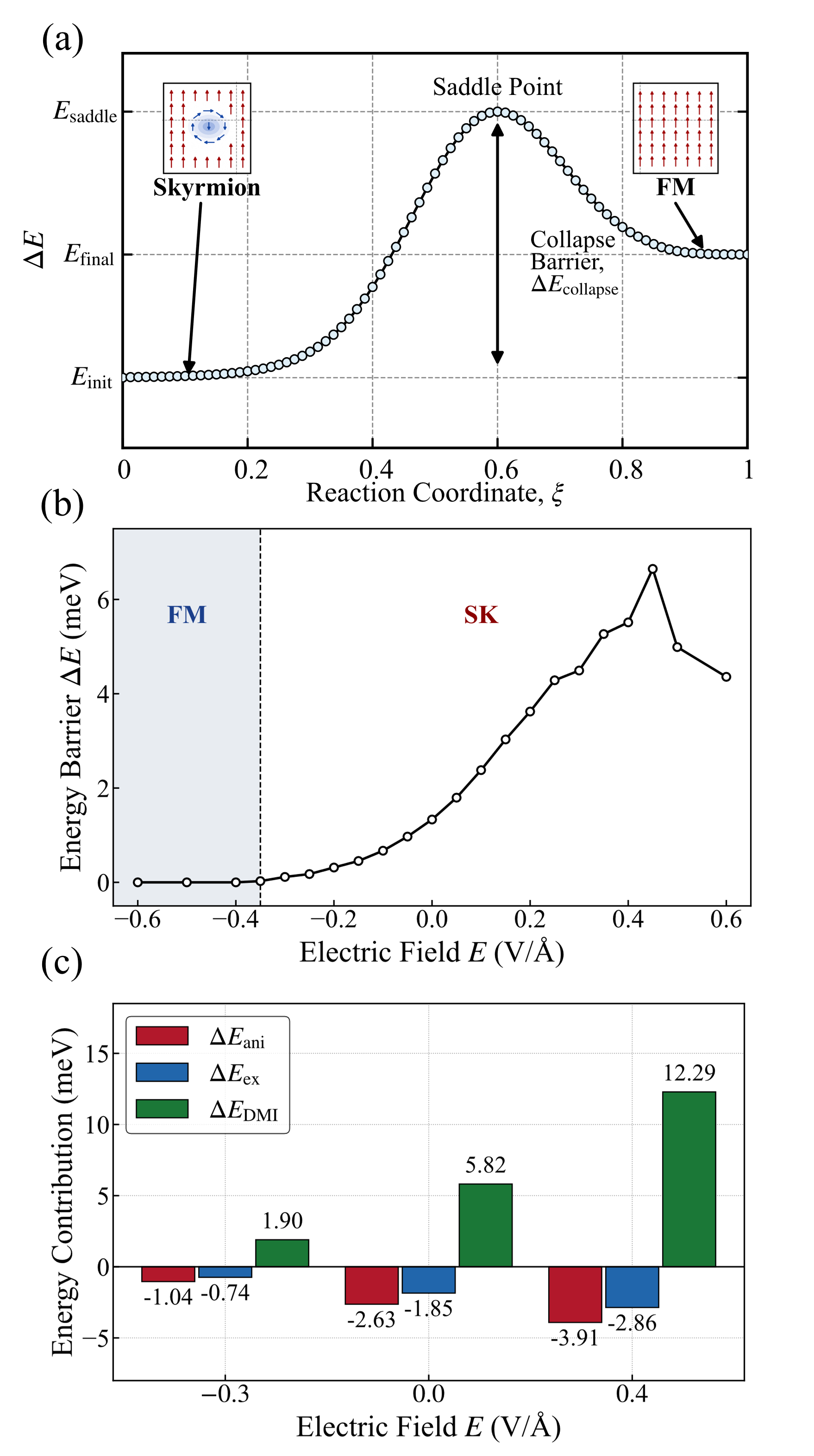}
\caption{(a) Collapse energy barrier of an isolated skyrmion versus external electric field, calculated via the GNEB method. (b) Minimum-energy paths connecting the skyrmion configuration to the uniform FM ground state. (c) Decomposition of the total energy barrier into Heisenberg exchange, MAE, and DMI contributions at zero electric field, revealing the microscopic physical origin of skyrmion stabilization.}
\label{fig:skyrmion_stability}
\end{figure}

The electric-field-induced magnetic phase transition from a collinear FM state to a skyrmionic state is accompanied by changes in the energetic stability of skyrmions. This stability can be quantitatively characterized by the collapse energy barrier of an isolated skyrmion embedded in an FM background. To gain more insights, the GNEB method is employed to calculate the collapse energy barrier of the skyrmion configuration in MBTSe under varying electric fields [Fig.~\hyperref[fig:skyrmion_stability]{5(a)}]. Being Consistent with the LLG simulation results, the overall stability of the skyrmion configuration is highly dependent on the applied field [Fig.~\hyperref[fig:skyrmion_stability]{5(b)}]. This indicates that increasing the electric field promotes the stabilization of skyrmions in the system. When the electric field exceeds $0.45$ V/\AA{}, the collapse energy barrier begins to decrease, indicating reduced skyrmion stability and in agreement with the high-field anomaly of the $\alpha$ parameter shown in Fig.~\hyperref[fig:magnetic_configuration]{4}. Consequently, the collapse energy barrier of a single skyrmion attains a distinct global maximum of approximately $6.5$ meV at exactly $0.45$ V/\AA{}. Given that an energy barrier in the range of 1 to 10 meV roughly corresponds to a thermal energy scale ($k_B T$) of 10 to 100 K, these values demonstrate that the skyrmion configuration can be robustly stabilized at cryogenic temperatures. Meanwhile, when the electric field reaches $-0.35$ V/\AA{}, the collapse energy barrier drops to nearly zero due to the influence of the strong out-of-plane MAE, returning the system to a stable FM state.

\begin{figure*}[t]
\centering
\includegraphics[width=0.83\textwidth]{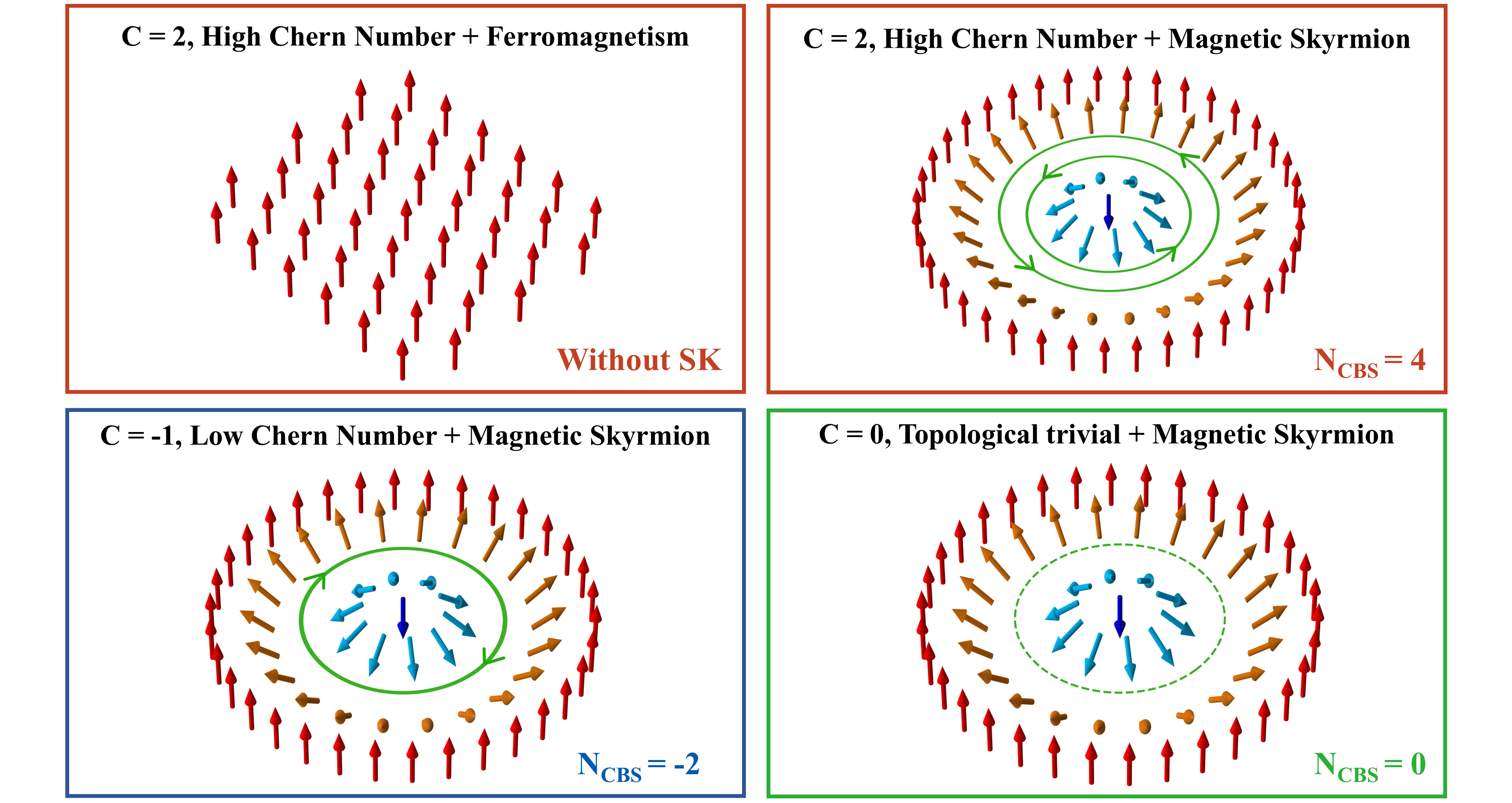}
\caption{Comprehensive multi-space topological phase diagram of Janus monolayer MBTSe under a perpendicular electric field. The system sequentially transitions across four distinct topological phases: ($C=2$, FM), ($C=2$, SK), ($C=-1$, SK), and ($C=0$, SK), featuring field-tunable CBSs at the skyrmion boundaries.}
\label{fig:phase_diagram}
\end{figure*}

To elucidate the microscopic mechanism by which the electric field controls the energy barrier, the total barrier is decomposed into distinct contributions from different magnetic interaction terms. Taking the zero electric field case as an example, compared to the highest energy magnetic state along the MEP, the exchange interaction contributes $1.84$ meV to the single skyrmion phase, the MAE contributes $2.63$ meV, and the DMI contributes $-5.82$ meV [Fig.~\hyperref[fig:skyrmion_stability]{5(c)}]. This implies that the DMI promotes the stabilization of the skyrmion, whereas the MAE and exchange interactions exert a suppressive effect. Among these, the competition between the DMI and MAE primarily governs the electric-field tunability of skyrmion stability. As the magnitude of the negative electric field increases, the MAE increases more rapidly than the DMI, rendering the skyrmion configuration progressively less stable.

\subsection{Overview of Topological Phases}

The simultaneous evolution of electronic band topology and magnetic spin textures in MBTSe gives rise to four tightly correlated multi-space topological phases. Specifically, as the perpendicular electric field is swept from negative to positive values, the system sequentially evolves through a purely k-space topological phase characterized by a high-Chern-number FM state ($C = 2$, FM), two hybrid topological phases featuring skyrmions with distinct electronic Chern numbers ($C = 2$, SK and $C = -1$, SK), and ultimately a purely real-space topological phase manifesting as a topologically trivial skyrmion state ($C = 0$, SK), as explicitly depicted in the comprehensive phase diagram [Fig.~\hyperref[fig:phase_diagram]{6}]. This electric-field-driven phase sequence provides a concise overview of how k-space and real-space topological orders can be jointly regulated in a single material platform.

Crucially, the coexistence of k-space QAH states and real-space skyrmion spin textures within the hybrid topological phases provides a rare physical platform for their direct interplay. This coexistence intrinsically accommodates the recently proposed multi-space topological excitation known as the RK joint topological skyrmion (RK-SK) \cite{li2022interplay,he2025coupling}. Fundamentally, an RK-SK is characterized by a topologically protected skyrmion encircled by dissipationless chiral boundary states (CBSs). The number and chirality of these surrounding CBSs are quantitatively determined by $N_{\mathrm{CBS}} = C_{\mathrm{inner}} - C_{\mathrm{outer}}$, where $C_{\mathrm{inner}}$ and $C_{\mathrm{outer}}$ denote the Chern numbers inside and outside the skyrmion, respectively. For an isolated N\'eel-type skyrmion with a spin-up core, the local magnetization reversal flips the sign of the inner Chern number (i.e. $C_{\mathrm{inner}} = -C_{\mathrm{outer}}$). Consequently, as the external electric field continuously tunes the background phase, $N_{\mathrm{CBS}}$ sequentially shifts from $4$ [in the ($C=2$, SK) phase] to $-2$ [in the ($C=-1$, SK) phase], and ultimately neutralizes to $0$ [in the ($C=0$, SK) phase]. Traditionally, conventional spintronic applications are predominantly designed around the creation and annihilation of skyrmions, which severely restricts device operations to simple binary states ($|1\rangle$ and $|0\rangle$). In stark contrast, this purely electric-field-driven, multi-step modulation of $N_{\mathrm{CBS}}$ introduces a fundamentally new manipulation paradigm which may help to increase the storage density in future spintronic devices.

\section{Conclusion}

In summary, the electric-field-induced manipulation of momentum-space band topology and real-space spin textures in Janus monolayer MBTSe has been systematically demonstrated. Under a varying perpendicular electric field, the system undergoes momentum-space transitions across topologically trivial ($C = 0$), low-Chern-number ($C = -1$), and high-Chern-number ($C = 2$) QAH states via bandgap closing and reopening. Concurrently, the modulation of intrinsic DMI and MAE drives real-space magnetic phase transitions spanning a uniform FM state, an ISK state, and a SK-SD coexistence phase. Crucially, sweeping the external electric field from negative to positive strengths unveils a completely reversible multi-space phase transition trajectory, evolving sequentially from a pure $k$-space topology, through a hybrid topological phase, to a pure real-space spin-texture topology. The realization not only broadens the fundamental understanding of multi-spatial topological crossovers but also paves a highly promising way for designing next-generation, multi-functional and dissipationless topological devices.

\begin{acknowledgments}

This work is supported by the National Key R\&D Program of China (Grant No. 2023YFA1406400).

\end{acknowledgments}

\section*{DATA AVAILABILITY}

The data that support the findings of this article are not publicly available upon publication because it is not technically feasible and/or the cost of preparing, depositing, and hosting the data would be prohibitive within the terms of this research project. The data are available from the authors upon reasonable request.


%\bibliography{apstemplate}

\end{document}